\documentclass[a4paper,11pt]{article}

\newcommand{\be}{\begin{eqnarray}}
\newcommand{\ee}{\end{eqnarray}}
\usepackage{jcappub} 
\usepackage{graphicx}
\usepackage{mathrsfs}

\title{Self-unitarization of New Higgs Inflation and compatibility with Planck and BICEP2 data}

\author[\dag]{Cristiano Germani}\author[\ddag]{, Yuki Watanabe}
\author[\dag]{and Nico Wintergerst}

\affiliation[\dag]{Arnold Sommerfeld Center, Ludwig-Maximilians-Universit\"at, Theresienstr. 37, 80333 M\"unchen, Germany}
\affiliation[\ddag]{Research Center for the Early Universe, University of Tokyo, Tokyo 113-0033, Japan}

\emailAdd{cristiano.germani@lmu.de}
\emailAdd{watanabe@resceu.s.u-tokyo.ac.jp}
\emailAdd{nico.wintergerst@physik.lmu.de}

\abstract{In this paper we show that the Germani-Kehagias model of Higgs inflation (or New Higgs Inflation), where the Higgs boson is kinetically non-minimally coupled to the Einstein tensor is in perfect compatibility with the latest Planck and BICEP2 data. Moreover, we show that the tension between the Planck and BICEP2 data can be relieved within the New Higgs inflation scenario by a negative running of the spectral index.
Regarding the unitarity of the model, we argue that it is unitary throughout the evolution of the Universe. Weak couplings in the Higgs-Higgs and Higgs-graviton sectors are provided by a large background dependent cut-off scale during inflation. In the same regime, the W and Z gauge bosons acquire a very large mass, thus decouple. On the other hand, if they are also non-minimally coupled to the Higgs boson, their effective masses can be enormously reduced. In this case, the W and Z bosons are no longer decoupled.
After inflation, the New Higgs model is well approximated by a quartic Galileon with a renormalizable potential. We argue that this can unitarily create the right conditions for inflation to eventually start.}
\subheader{RESCEU-8/14\\LMU-ASC 13/14}
\begin{document}
\maketitle
\flushbottom
\section{Higgs boson as the inflaton}
Recently, there have been three extremely important discoveries:
\begin{itemize}
\item The Standard Model (SM) Higgs boson has been observed in the Large Hadron Collider (LHC) of Geneva \cite{lhc}. In the same experiment, no new particles, beyond the SM, have been discovered so far.
\item The European Planck satellite has measured, with unprecedented precision, the primordial spectrum of scalar (temperature) perturbations, showing no trace of non-gaussianities and isocurvature perturbations \cite{planck}.
\item The U.S.A. BICEP2 experiment has measured the polarization of the B-modes in the Cosmic Microwave Background (CMB), thus providing the first evidence for primordial gravitational waves \cite{bicep}.
\end{itemize}
If the results of Planck and BICEP2 are confirmed, they provide striking evidence for the existence of an inflationary stage in our Universe. 
 
The null observation of isocurvature and non-gaussian modes in the CMB point to the simplest model of inflation, the one generated by a single scalar field, the inflaton. At the same time, the large spectrum of gravitational waves, as now apparently measured by BICEP2, singles out chaotic type models of inflation \cite{linde}, where the (canonically normalized) inflaton ranges over trans-Planckian values.  Historically, this fact has always been matter of debate. Naively one would indeed expect Planck suppressed operators (from quantum gravity) correcting the inflationary potential. 
However, these corrections will be suppressed if they only appear as expansions in powers of the potential itself, which is always way below the Planck scale \cite{linden} (see also \cite{riottoke}).

With the LHC discovery, it is tempting to consider the very minimal scenario where the Higgs boson not only accounts for the masses of the SM particles, but also for inflation. 

In absence of gravity, the Higgs boson Lagrangian is
\be
{\cal L}_{\cal H}=-{\cal D}_\mu {\cal H^\dag}{\cal D}^\mu {\cal H}-\lambda\left({\cal H}^\dag{\cal H}-v^2\right)^2\ ,
\ee
where ${\cal D}_\mu=\partial_\mu-i g W^a_\mu\tau^a-i\frac{g'}{2} B_\mu$ is the covariant derivative related to the $SU(2)$ gauge bosons $W_\mu^a$ with generator $\tau^a$ and the $U(1)_Y$ gauge boson $B_\mu$. The Higgs boson is a complex doublet of SU(2) (charged under $U(1)_{Y}$). The scale $v\sim 246$ GeV is very low compared to the Higgs background during inflation, we 
can therefore safely neglect it.

Forgetting for a moment the contributions of the gauge sectors of the SM, and focusing only on the radial part of the Higgs boson $\phi\sim \sqrt{2{\cal H}^\dag{\cal H}}$, we reduce to the following Lagrangian
\be
{\cal L}_{\phi}=-\frac{1}{2}\partial_\mu \phi \partial^\mu \phi-\frac{\lambda}{4}\phi^4\ .
\ee
Defining the slow-roll parameters that parametrize how close the the system is evolving to a de Sitter background  \cite{mukhanov}
\be\label{epsilongr}
\epsilon_{V}\equiv\frac{V'^2M_p^2}{2V^2}\ \mbox{and}\ \eta_{V}\equiv\frac{V'' M_p^2}{V}\ ,
\ee
where we considered a generic potential $V$ for the inflaton ($\phi$) and $V'=dV/d\phi$, one finds that the power spectrum of primordial perturbations, quantum mechanically generated by the inflaton during inflation, is
\be
{\cal P}\simeq \frac{H^2}{8\pi^2\epsilon_{V} M_p^2}\ ,
\ee 
and the spectral index
\be
n_s=1-6\epsilon_{V}+2\eta_{V}\ .
\ee
Planck data constrains ${\cal P}\sim 10^{-9}$ and $n_s\sim 0.96$. This requires an extremely small value for $\lambda\sim 10^{-12}$ which would imply an extreme fine tuning for the top quark mass if running of coupling constants in SM is taken into account \cite{running}.

This fine tuning can be alleviated by considering a non-SM Higgs boson potential, interpolating from the low energy quartic SM coupling, to a high-energy flatter potential. 
However, in this case the pure Higgs sector of the SM must pass through a strong coupling below the inflationary and above the LHC scales. This automatically requires a UV completion by other degrees of freedom at intermediate scale (see for example \cite{riotto}). Therefore, in those cases, this ``Higgs" potential cannot be directly connected to the SM Higgs potential (for latest examples see \cite{shaposhnikov, fumi}).

It is interesting to mention that one of the most popular example of Higgs inflation of this class of models, the one of \cite{shaposhnikov}, is now ruled out by BICEP2. In the Higgs inflation of \cite{shaposhnikov} a conformal coupling of the Higgs boson to the Ricci scalar of the form $\xi\phi^2 R$ effectively introduces an exponentially flat potential which now predicts a too low gravitational wave spectrum. 

An alternative to consider a new potential for the Higgs boson has been introduced in \cite{new}. In the New Higgs Inflationary scenario of \cite{new}, the Higgs boson kinetic term is (uniquely) non-minimally coupled to the Einstein tensor as follows
\be
{\cal L}_{kin}=-\frac{1}{2}\left(g^{\alpha\beta}-\frac{G^{\alpha\beta}}{M^2}\right)\partial_\alpha\phi\partial_\beta\phi\ .
\ee
The above coupling does not introduce any new degrees of freedom other than the graviton and the Higgs boson and, in particular, no higher derivative terms. Interesting enough, this kind of non-minimal coupling can also be obtained in supergravity \cite{fotis, fotis2}. Finally, the mass scale $M$ must be determined experimentally.

For a sufficiently low scale $M$, but much bigger than LHC scales \cite{alexpert, yuki}, the non-minimal coupling of the Higgs boson to gravity introduces an enhanced friction acting on the scalar $\phi$ making $\phi$ slowly rolling even for large (e.g. $\lambda\sim 0.1$) values \cite{yuki,somehow}. In the same regime, the New Higgs inflation scenario is able to match observational data (for previous results in compatibility with WMAP \cite{wmap}, see \cite{yuki}). The interesting point is that, since in this case the quartic Higgs coupling can assume any value, the fine tuning of the top quark mass afflicting the General Relativistic (GR) case is removed. 

In addition, the New Higgs inflation, being a chaotic-like inflationary model, conversely to previous attempts \cite{shaposhnikov} also produces a gravitational wave spectrum which is simultaneously compatible with Planck and BICEP2, as we shall show in the following.

\section{Fitting Planck and BICEP2 data with New Higgs Inflation}

As announced before, we shall consider a generic inflationary Lagrangian of gravity and a scalar field kinetically coupled to gravity in the following form:
\be\label{eq:lagrangian_full}
{\cal L}=\sqrt{-g}\left[\frac12 M_p^2R -\frac12\left(g^{\mu\nu} - \frac{G^{\mu\nu}}{M^2}\right)\partial_{\mu}\phi\partial_{\nu}\phi  -V(\phi)\right],
\ee
where the sign of the kinetic gravitational coupling was chosen to avoid appearance of ghost instabilities in perturbations \cite{new, alexpert, yuki}.\footnote{It is known that this model is equivalent to another form: \cite{Kobayashi:2011nu}
\be
{\cal L}=\sqrt{-g}\left[\frac12 M_p^2R -\frac12 g^{\mu\nu}\partial_{\mu}\phi\partial_{\nu}\phi + \frac{1}{2M^2} \left(-\frac{1}{2}(\partial_{\mu}\phi)^2R +(\Box \phi)^2-(\nabla_{\mu}\nabla_{\nu}\phi)^2 \right) -V(\phi)\right],\nonumber
\ee
where integration by parts has been done. See also \cite{Kamada} for studies of Higgs inflation in the context of the most general single-field model.}
In the Arnowitt-Deser-Misner (ADM) formalism with a metric $ds^2=-N^2dt^2+h_{ij}(N^i dt+dx^i)(N^j dt+dx^j)$, all geometrical quantities are described by defining a spatial covariant derivative $D_i$, a three-dimensional curvature scalar ${}^{(3)}R$ and extrinsic curvature $K_{ij}=E_{ij}/N$ all constructed from $h_{ij}$. The Lagrangian~(\ref{eq:lagrangian_full}) then reads
\be\label{eq:lagrangian_adm}
{\cal L}&=&\sqrt{h}\frac{M_p^2}{2}\left[ {}^{(3)}R\left(N+\frac{\dot\phi^2}{2NM^2M_p^2}\right) + (E_{ij}E^{ij}-E^2)\left(\frac1N - \frac{\dot\phi^2}{2N^3M^2M_p^2}\right) + \frac{\dot\phi^2}{NM_p^2}-\frac{2NV}{M_p^2}\right],\nonumber\\
E_{ij}&=&\frac12 (\dot{h}_{ij}-D_iN_j-D_jN_i), \quad
E=h^{ij}E_{ij},
\ee
where the unitary gauge $\phi(x,t) = \phi(t)$ has been chosen.
We further assume an almost flat Friedmann universe with a metric
\be
 N=1+\alpha,\quad 
 N_i=\partial_i\beta,\quad 
 h_{ij}=a^2e^{2\zeta}(\delta_{ij}+\gamma_{ij}+\gamma_{il}\gamma_{lj}/2)
 \ee
 to second order, where $\zeta$ is the curvature perturbation and $\gamma_{ij}$ are gravitational waves with transverse-traceless conditions $D^i\gamma_{ij}=0$ and $h^{ij}\gamma_{ij}=0$.
 
Varying the Lagrangian~(\ref{eq:lagrangian_adm}) with respect to the lapse $N$, we find the Hamiltonian constraint equation
\be\label{eq:hamiltonian_constraint}
{}^{(3)}R\left( N^2-\frac{\dot\phi^2}{2M^2M_p^2}\right)-(E_{ij}E^{ij}-E^2)\left(1-\frac{3\dot\phi^2}{2N^2M^2M_p^2}\right)- \frac{\dot\phi^2}{M_p^2}-\frac{2N^2V}{M_p^2}=0.
\ee
This equation to zeroth order gives the Friedmann equation
\be\label{eq:friedmann}
H^2 = \frac{1}{3M_p^2}\left[ \frac{\dot\phi^2}{2}\left(1+\frac{9H^2}{M^2}\right)+V\right]\ ,
\ee
where we have used $\bar{E}_{ij}\bar{E}^{ij}-\bar{E}^2=-6H^2$ and ${}^{(3)}\bar{R}=0$.
Varying the Lagrangian~(\ref{eq:lagrangian_adm}) with respect to $\phi(t)$, we find the background equation of motion for $\phi$ as
\be\label{eq:phi}
\frac{1}{a^3}\frac{d}{dt}\left[a^3\dot\phi\left(1+\frac{3H^2}{M^2}\right) \right]+V'=0.
\ee
Varying the Lagrangian~(\ref{eq:lagrangian_adm}) with respect to $a(t)$ and combining with (\ref{eq:friedmann}), we get another useful background relation
\be\label{eq:raychauduri}
-\frac{\dot{H}}{H^2}\left( 1-\frac{\dot\phi^2}{2M^2M_p^2}\right)=\frac{\dot\phi^2}{2H^2M_p^2}\left(1+\frac{3H^2}{M^2}\right)-\frac{\ddot\phi\dot\phi}{HM^2M_p^2}.
\ee

Varying the Lagrangian~(\ref{eq:lagrangian_adm}) with respect to the shift $N_j$, we find the momentum constraint equation
\be
D^i\left[\left(\frac1N - \frac{\dot\phi^2}{2N^3M^2M_p^2}\right)(E_{ij}-h_{ij}E) \right]=0,
\ee
whose solution to first order is given by \cite{yuki}
\be\label{eq:lapse}
\alpha = \frac{\Gamma}{H}\dot\zeta,\quad
\Gamma \equiv \frac{1-\frac{\dot\phi^2}{2M^2M_p^2}}{1-\frac{3\dot\phi^2}{2M^2M_p^2}}.
\ee
To first order in (\ref{eq:hamiltonian_constraint}) we get 
\be
\delta {}^{(3)}R\left(1-\frac{\dot\phi^2}{2M^2M_p^2}\right) &-& \delta( E_{ij}E^{ij}-E^2) \left(1-\frac{\dot\phi^2}{2M^2M_p^2}\right)+\frac{18H^2\dot\phi^2\alpha}{M^2M_p^2} -\frac{4V\alpha}{M_p^2}=0,\nonumber\\
\delta {}^{(3)}R &=&-4D^2\zeta,\quad
\delta( E_{ij}E^{ij}-E^2)=-12H\dot\zeta+4HD^2\beta,
\ee
whose solution is given by \cite{yuki}
\be\label{eq:shift}
\beta=-\frac{\Gamma}{H}\zeta +\chi,\quad
\partial_i^2\chi=\frac{a^2\Gamma^2\Sigma}{H^2\left(1-\frac{\dot\phi^2}{2M^2M_p^2}\right)}\dot\zeta,\quad
\Sigma \equiv \frac{\dot\phi^2}{2M_p^2}\left[1+\frac{3H^2(1+\frac{3\dot\phi^2}{M^2M_p^2})}{M^2(1-\frac{\dot\phi^2}{2M^2M_p^2})}\right],
\ee
where the Friedmann equation (\ref{eq:friedmann}) and the lapse (\ref{eq:lapse}) have been used to get the shift.

To second order in the Lagrangian~(\ref{eq:lagrangian_adm}), scalar and tensor modes decouple. We obtain the quadratic Lagrangian in $\zeta$ after a few integration by parts:
\be
{\cal L}_{\zeta^2}&=&a^3M_p^2 \left[ \frac{\Gamma^2\Sigma}{H^2}\dot\zeta^2-\frac{\epsilon_s}{a^2}(\partial_i\zeta)^2\right],\nonumber\\
\epsilon_s &\equiv& \frac1a\frac{d}{dt}\left[\frac{a\Gamma}{H}\left(1-\frac{\dot\phi^2}{2M^2M_p^2}\right) \right]-\left(1+\frac{\dot\phi^2}{2M^2M_p^2}\right),
\ee
where we have used background and first-order equations~(\ref{eq:friedmann}), (\ref{eq:raychauduri}), (\ref{eq:lapse}) and (\ref{eq:shift}). From the Friedmann equation (\ref{eq:friedmann}), $0<\dot\phi^2/(M^2M_p^2) < 2/3$, and thus the positive-definite coefficient of $\dot\zeta^2$
 is guaranteed in the quadratic action, i.e. $\Gamma^2\Sigma/H^2>0$; this indicates that the curvature perturbation cannot be ghost-like in the Friedmann background with $V \ge 0$.
Note that $\zeta$ is ill-defined when $\dot\phi=0$, and we have to choose another gauge (or variable) to describe perturbations. We will stretch this point later on in connection to the unitarity issues of our theory.

A gradient instability can be avoided if $\epsilon_s >0$, which is true as long as
\be
\epsilon \equiv -\frac{\dot H}{H^2} < \frac{11}{6}+\frac{5M^2}{6H^2}+\frac{M^4}{12H^4},
\ee 
where we have used (\ref{eq:raychauduri}).

The coefficients of the quadratic Lagrangian, $\Gamma^2\Sigma/H^2$ and $\epsilon_s$, are governed by background equations~(\ref{eq:friedmann}), (\ref{eq:phi}) and (\ref{eq:raychauduri}). During slow rolling of $\phi$, we obtain a quasi-DeSitter background universe that is described by
\be\label{eq:slow-roll}
H^2 \simeq \frac{V}{3M_p^2},\quad
\dot\phi \simeq -\frac{V'}{3H\left(1+\frac{3H^2}{M^2} \right)},\quad
\epsilon=-\frac{\dot H}{H^2}\simeq \frac{\dot\phi^2}{2H^2M_p^2}\left(1+\frac{3H^2}{M^2}\right),
\ee
where $\epsilon \ll 1$ and $\delta \equiv \ddot\phi/(H\dot\phi)\ll 1$.\footnote{In terms of the potential, we can express the slow-roll parameters as
\be
\epsilon \simeq \frac{V'^2M_p^2}{2V^2\left( 1+ \frac{V}{M^2M_p^2}\right)},\quad
\delta \simeq -\frac{V''M_p^2}{V\left( 1+ \frac{V}{M^2M_p^2}\right)}+\epsilon+2\epsilon\frac{V}{M^2M_p^2\left(1+ \frac{V}{M^2M_p^2}\right)}.
\nonumber
\ee
}
Under this approximation, relations such that $\Gamma^2\Sigma/H^2 \simeq \epsilon_s \simeq \epsilon$ hold.

In order to evaluate the primordial power spectrum from inflation, we canonically quantize $\zeta$ by using conformal time $\tau$ and the Mukhanov-Sasaki variable $v = z\zeta$ \cite{mukhanov}, where $z = aM_p\Gamma\sqrt{2\Sigma}/H\simeq aM_p\sqrt{2\epsilon}$. Since background quantities ($\dot\phi$, $H$ and $\Sigma$) are changing slowly during inflation, a non-decaying solution for $\zeta_k$ can be found on super-horizon scales. These modes stay constant until the horizon re-entry. The power spectrum of primordial curvature perturbations is then given by \cite{yuki}
\be\label{eq:scalar-spectrum}
{\cal P}_{\zeta}=\frac{k^3}{2\pi^2}|\zeta_k|^2 \simeq \frac{H^2}{8\pi^2\epsilon_s c_s M_p^2},\quad
c_s^2 = \frac{H^2\epsilon_s}{\Gamma^2\Sigma},
\ee
where the Hubble scale has been evaluated at the sound horizon exit, $c_sk= aH$. Note that in the limit $3H^2 \ll M^2$ (conventional general relativity limit), $c_s^2 \simeq 1$ and in the other limit $3H^2 \gg M^2$ (gravitationally enhanced friction limit), $c_s^2 \simeq 1-8\epsilon/3$.
Although these limits have no impact on the amplitude of curvature perturbation, they do on its spectral index and change theoretical predictions.
The scalar spectral index is given by 
\be\label{eq:scalar-tilt}
n_s-1 \simeq -2\epsilon-\frac{\dot{\epsilon}_s}{\epsilon_sH}=-4\epsilon\left(1-\frac{3H^2}{2M^2(1+\frac{3H^2}{M^2})}\right) -2\delta.
\ee

Similarly, gravitational waves can be canonically quantized and described by the Mukhanov-Sasaki equation with $v_t=z_t\gamma_{\lambda}$, $z_t=aM_p\sqrt{1-\dot\phi^2/(2M^2M_p^2)}\simeq aM_p$ and $c_t^2=[1+\dot\phi^2/(2M^2M_p^2)]/[1-\dot\phi^2/(2M^2M_p^2)] \simeq 1$ as in the scalar modes. The tensor power spectrum is given by \cite{yuki}
\be\label{eq:tensor-spectrum}
{\cal P}_{\gamma}=\frac{k^3}{2\pi^2}\sum_{\lambda=+,\times}|\gamma_{\lambda}(k)e_{ij}^{\lambda}(k)|^2
 \simeq \frac{2H^2}{\pi^2c_tM_p^2(1+\frac{\dot\phi^2}{2M^2M_p^2})},
\ee
which has been evaluated at the tensor horizon exit, $c_t k= aH$.
The tensor spectral index is given by 
\be
n_t \simeq -2\epsilon.
\ee
From (\ref{eq:scalar-spectrum}) and (\ref{eq:tensor-spectrum}), the tensor-to-scalar-ratio is given by
\be
r \simeq 16\epsilon,
\ee
which yields a consistency relation
\be
r\simeq-8n_t.
\ee

\begin{figure}[tb]
  \centering
    \includegraphics[width=0.8\textwidth]{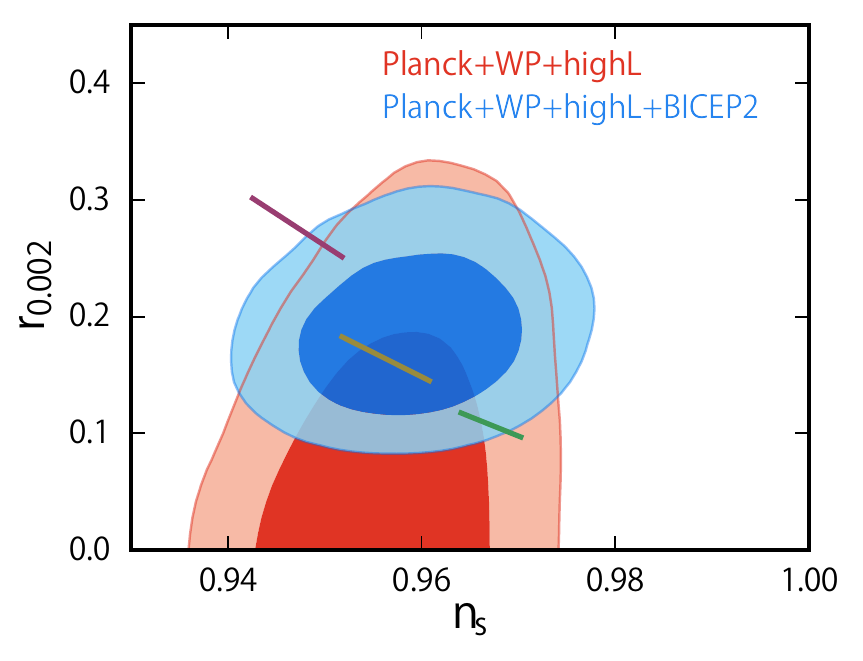}
\caption{The contours show $1\sigma$ and $2\sigma$ constraints on $n_s$ and $r$, taken from figure~13 of \cite{bicep}. The lines show predictions of New Higgs inflation with $N_*=50-60$ (left to right end of each line) and different values of a theoretical parameter, $\lambda M_p^2/M^2$. The highest line (purple) corresponds to the GR limit, the lowest (green) to the high friction limit $H^2/M^2\simeq 2\times 10^5$ (for $\lambda =0.1$ and ${\cal P}_{\zeta}=2\times 10^{-9}$) and the middle (yellow) to the best value of $\lambda M_p^2/M^2=2.2\times 10^{-4}$. The latter, {\it does not} describe an inflating background in the high friction limit. It is numerically found for $H^2/M^2\simeq 1.2$, here $\lambda\simeq 3.9\times10^{-13}$ and $N_*=50$.}
\label{fig}
\end{figure}

In figure \ref{fig}, we numerically solve the slow-roll equations \eqref{eq:slow-roll} and show a sample of possible matchings of the New Higgs Inflation predictions with the combined Planck and BICEP2 data sets chosen by the BICEP2 team \cite{bicep}. In particular one finds that, for any value of $\lambda$, the quartic Higgs coupling, there is always a value for $M$ such to fit the combined Planck and BICEP2 data within 2-$\sigma$. The central value in figure \ref{fig}, fitting the data within 1-$\sigma$, is for a hybrid case in which the enhanced friction due to the non-minimal coupling $\sim H\times \frac{3 H^2}{M^2}$, is comparable to the GR friction $\sim H$. In this case though, the Higgs quartic coupling has to run to order $10^{-13}$ at the inflationary scale. Whether this is the case or not, can only be known after a better measure of the top-quark mass and Yukawa coupling in collider experiments.

\subsection{Relaxing the tension between Planck and BICEP2}

The tensor-to-scalar-ratio $r$ measured by BICEP2 is in mild tension with the best value of $r$ found by the Planck team. As it has been suggested already by the BICEP2 team, a possible resolution to this tension is to relax some of the assumption used in the data analysis performed by the Planck team. 

One obvious possibility is to allow for a running of the scalar spectral index. Although other explanations of this tension are also possible, such as a better measure of the optic depth, it is intriguing to see that the New Higgs inflationary model can provide a non-trivial negative running of the spectral index relaxing the tension between Planck and BICEP2. We will be very quantitative here.

In gravitationally enhanced friction (GEF) limit $3H^2\gg M^2$, we get the scalar spectral index from (\ref{eq:scalar-tilt})
\be
n_s-1 = -2\epsilon -2\delta = (-8\epsilon_{V}+2\eta_{V})\frac{M^2}{3H^2},
\ee
which is different from the GR one, $n_s-1=-4\epsilon-2\delta = -6\epsilon_{V}+2\eta_{V}$ in the other limit; the difference indeed makes the $\lambda\phi^4/4$ potential (or other non-flat potentials \cite{yuki, Tsujikawa}) compatible with the Planck and BICEP2 observations since a relevant quantity is not $\epsilon_V$ but $\epsilon=\epsilon_VM^2/(3H^2)$ (in the GEF limit). The running of the spectral index in the GEF limit is given by
\be
\left.\frac{dn_s}{d\ln{k}}\right|_{c_sk=aH}=-6\epsilon\delta -2\delta\delta'+2\delta^2 = ( 24\epsilon_V\eta_V-48\epsilon_V^2-2\xi_V^2 )\frac{M^4}{9H^4}, 
\ee
where $\delta'\equiv \dddot\phi/(\ddot\phi H)$ and $\xi_V^2\equiv V'''V'M_p^4/V^2$. 

According to the above equations, the New Higgs inflation predicts
\be
n_s&-&1 = -5 \epsilon,\quad
r=16\epsilon,\quad
\frac{dn_s}{d\ln k}=-15\epsilon^2,\nonumber\\
\frac{\phi_*}{M_p} &=& 0.037 \left(\frac{{\cal P}_\zeta}{2\times 10^{-9}}\right)^{1/4}\left(\frac{\epsilon}{\lambda}\right)^{1/4},\quad
\frac{M}{M_p}= 9.0\times 10^{-6}\left(\frac{{\cal P}_\zeta}{2\times 10^{-9}}\right)^{3/4}\frac{\epsilon^{5/4}}{\lambda^{1/4}},\nonumber\\
\frac{H}{M_p} &=& 4.0\times 10^{-4}\left(\frac{{\cal P}_\zeta}{2\times 10^{-9}}\right)^{1/2} \sqrt{\epsilon},\quad
\epsilon =\frac{1}{3N_*+1},
\ee 
where $N_*$ is the number of e-folds at the CMB scale.

In Planck Collaboration (2013) XXII~\cite{planck}, the constraints on the parameters for $\Lambda {\rm CDM}+r+dn_s/d\ln{k}$ model are given from Planck combined with other data sets. In the data set Planck+WP+BAO, 
\be
n_s = 0.9607\pm 0.0126,\quad r<0.25 \quad \text{and}\quad 
\frac{dn_s}{d\ln{k} }=-0.021^{+0.024}_{-0.020} \nonumber
\ee
all at $2\sigma$. In the data set Planck+WP+high-$\ell$, 
\be
n_s = 0.9570\pm 0.0150,\quad r<0.23\quad 
\text{and}\quad 
\frac{dn_s}{d\ln{k} }=-0.022^{+0.022}_{-0.020} \nonumber 
\ee
all at $2\sigma$, though BICEP2 team \cite{bicep} cited a slightly different upper bound for the running as $dn_s/d\ln{k}=-0.022\pm 0.020$. If we allow the non-vanishing running, the New Higgs inflation's predictions can be well within the constraints from Planck and BICEP2. The New Higgs inflation generically predicts $dn_s/d\ln{k}\sim -10^3$ slightly larger than that from other simple chaotic inflation models.
  
Specifically we have the following predictions compatible with the Planck and BICEP2 data
\be
n_s= 0.95,\quad
r= 0.16,\quad
\frac{dn_s}{d\ln{k}}=-0.0015,
\ee
with $N_*=33$.  Compared with minimal gravity (GR), the reported $r$ is a few times smaller and the running is a few times larger, for the same $n_s$.  

\section{Unitarity issues: inflationary scale}

In this section we complete the previous analysis of \cite{new} by showing that the New Higgs inflation is weakly coupled during the whole inflationary evolution.

We will initially focus on the scalar sector of the theory. The introduction of a non-minimal kinetic term for the Higgs boson introduces a new non-renormalizable interaction to the SM
\be
{\cal L}_{\rm nr}=\frac{1}{2}\frac{G^{\alpha\beta}}{M^2}\partial_\alpha\phi\partial_\beta\phi\ .
\ee 
The question is then what is the tree-level unitarity violating scale of this system, and whether our inflating background is safely below this scale, as to be able to trust our semiclassical calculations.

During inflation, and in the high friction regime, the perturbed Lagrangian around the inflating background, up to cubic order, is
\be\label{Lp}
{\cal L}_{\delta\phi,h}&=&-\frac{M_p^2}{2}h^{\alpha\beta}{\cal E}_{\alpha\beta}-\frac{3H^2}{2M^2}\partial_\mu\delta\phi\partial^\mu\delta\phi+\frac{{\cal E}^{\alpha t}}{M^2}\dot\phi_0\partial_\alpha\delta\phi+\frac{{\cal E}^{\alpha\beta}}{2M^2}\partial_\alpha\delta\phi\partial_\beta\delta\phi+\cr
&+&h^{\alpha t}\dot\phi_0\partial_\alpha\delta\phi+\frac{1}{2}h^{\alpha\beta}\partial_\alpha\delta\phi\partial_\beta\delta\phi+\ldots\ .
\ee
In the above Lagrangian the Higgs boson has been split as $\phi=\phi_0+\delta\phi$ where $\phi_0$ is the background value and $\delta\phi$ the perturbation. The metric has been expanded as $g_{\alpha\beta}=g^{(0)}_{\alpha\beta}+h_{\alpha\beta}$ where $g^{(0)}_{\alpha\beta}$ is the background metric and $h_{\alpha\beta}$ is the metric perturbation. Finally, ${\cal E}_{\alpha\beta}$ is the linearized Einstein tensor on the inflating background.

By looking at the Lagrangian \eqref{Lp}, it is clear that neither the graviton nor the Higgs boson are canonically normalized, and in fact also mix. 
The canonical normalizations of these fields are
\be
\bar h_{\alpha\beta}&=&M_p h_{\alpha\beta}\cr \nonumber
\bar\phi&=& \frac{\sqrt{3}H}{M}\delta\phi\ .
\ee 
Thus we have
\be
{\cal L}_{\delta\phi,h}&=&-\frac{1}{2}\bar h^{\alpha\beta}{\cal E}(\bar h)_{\alpha\beta}-\frac{1}{2}\partial_\mu\bar\phi\partial^\mu\bar\phi+\frac{{\cal E}(\bar h)^{\alpha\beta}}{2H^2 M_p}\partial_\alpha\bar\phi\partial_\beta\bar\phi+\text{mixings}\ldots\ .
\ee
It would then seem that the strong coupling scale of this theory is $\Lambda_H\sim\left(H^2 M_p\right)^{1/3}\ll M_p$.

However, as we shall see, the scale $\Lambda_H$ will be removed by diagonalization of the scalar-graviton system. The easy way to do that is to use the the diffeomorphism invariance of the theory and go in what is usually called the unitary gauge \cite{maldacena}.

Allowing the freedom of time reparameterization, in an inflating background we have
\be
\phi(t+\delta t,x)=\phi_0(t)+\delta\phi+\dot\phi_0\delta t+\ldots\ .
\ee	
Now it is clear that one can always reabsorb the scalar fluctuations by choosing 
\be
\delta t=-\frac{\delta\phi}{\dot\phi_0}\ .
\ee
This is a very well known fact in cosmological perturbations (see for example \cite{maldacena} for a heavy use of this).

In this gauge, we see that all interactions of the scalar $\bar \phi$ to the longitudinal graviton are gauged away and the strong coupling scale of the theory becomes $M_p$. This is of course true only at leading order in slow-roll. As shown by the explicit computations of \cite{yuki}, the scalar (longitudinal graviton) interactions during inflation are indeed governed by the strong coupling scale $M_p/\sqrt{\epsilon}$. The tensor (transverse graviton) interactions provide the strong coupling scale
\be
\Lambda=M_p\sqrt{1-\frac{\dot\phi^2}{2M^2 M_p^2}}\ ,
\ee
which, at leading order in slow roll, is $\Lambda\simeq M_p$, as mentioned before.  

It is curious to see that in a Friedmann background the strong coupling scale is bounded by $M_p\sqrt{\frac{2}{3}}$. This can be seen from the Friedmann equations \eqref{eq:friedmann}
\be
H^2=\frac{\dot\phi^2}{6 M_p^2}\left(1+\frac{9 H^2}{M^2}\right)+\frac{V}{3M_p^2}\ .
\ee
For a positive definite potential we have the following bound
\be
\frac{3\dot\phi^2}{2M^2 M_p^2}\leq 1\ ,
\ee
which sets the smallest strong coupling scale to be $M_p\sqrt{\frac{2}{3}}$.

\subsection{The gauge bosons}

If the Higgs boson was just a real scalar field, the discussion before about the strong scale would be exhaustive. However, the Higgs boson is {\it not} a real scalar field. The Higgs boson being a complex doublet of the SU(2)  and charged under $U(1)_{Y}$, will couple to the SM gauge bosons as well. In order to keep gauge invariance, the coupling must be of the form
\be\label{gauge}
{\cal L}_{gauge}=-\left(g^{\alpha\beta}-\frac{G^{\alpha\beta}}{M^2}\right){\cal D}_\alpha{\cal H}^\dag {\cal D}_\beta{\cal H}\ .
\ee
On a background for the Higgs, the gauge fields obtain masses proportional to the Higgs vev via the Higgs mechanism. In the unitary gauge, in which the Higgs boson is only parameterized by $\phi$, the Goldstone bosons are eaten up by the gauge vectors. During inflation it is then easy to see that the gauge vectors W and Z will become heavy with a mass $m_{W,Z}=g_{W, Z}\phi_0\frac{\sqrt{3}H}{M}> M_p$.\footnote{
This is an entirely gauge invariant statement. To emphasize this point, let us discuss the simple example of a $U(1)$-symmetry, spontaneously broken through the radial background component $\phi_0$ (which we take slowly varying as in the slow roll evolution we are interested in) of a complex scalar $\Phi\equiv \phi e^{i \theta}/\sqrt2$. In this case, the kinetic term of the gauge-Higgs sector (including the non-minimal coupling of the scalar to gravity) reads
\begin{equation}
{\cal L}_K = -\frac{1}{4}F_{\mu\nu}^2 - \frac{1}{2} \left(g^{\mu\nu} - \frac{G^{\mu\nu}}{M^2}\right)\left(\partial_\mu \delta\phi\,\partial_\nu \delta\phi+\partial_\mu \vartheta\,\partial_\nu \vartheta + g^2 \phi_0^2 A_\mu A_\nu + 2 g \phi_0 A_\mu \partial_\nu \vartheta\right)\,,\nonumber
\end{equation}
where we have neglected derivatives of $\phi_0$, as slow rolling. In the kinetic action written above $A_\mu$ is the gauge boson, $F_{\mu\nu}$ the corresponding field strength, $\vartheta\simeq \phi_0\theta$ the (approximate) Goldstone boson and we have expanded the Higgs boson on the background, i.e. $\phi = \phi_0+\delta\phi$.

Due to the presence of the mixing between $A_{\mu}$ and $\vartheta$, controlled by the mass $m \sim \sqrt{1 + 3H^2/M^2} g \phi_0$, the Goldstone bosons are only good degrees of freedom at energies much larger than the mass. At lower energies, one should instead consider the \emph{gauge invariant} combination $W_\mu \equiv A_\mu + \frac{1}{m}\partial_\mu \vartheta$. In terms of $W_\mu$ and its field strength $W_{\mu\nu}$, the {\it gauge invariant} Lagrangian reads
\begin{equation}
{\cal L}_K = -\frac{1}{4}W_{\mu\nu}^2 - \frac{1}{2} \left(g^{\mu\nu} - \frac{G^{\mu\nu}}{M^2}\right) g^2 \phi_0^2 W_\mu W_\nu  - \frac{1}{2} \left(g^{\mu\nu} - \frac{G^{\mu\nu}}{M^2}\right)\partial_{\mu}\delta\phi\partial_{\nu}\delta\phi \,,\nonumber
\end{equation}
which takes the same form of the action written in unitary gauge. This is the essence of the St\"uckelberg decomposition of a massive vector field.

In the case of a spontaneously broken $SU(2) \times U(1)$-theory, the issue is technically slightly more involved due to the nonlinearity of the gauge transformations, but physically equivalent.}  In any theory of gravity, particles with masses larger than $M_p$ can be considered as Black Holes \cite{sarah} (note that Black Holes in this theory are the same as in GR \cite{luca}). In this case, one could think to abandon the elementary particle description of the gauge bosons and integrate them out as if they were extended objects, i.e. Black Holes. In this case, practically, we could forget about this sector, as mentioned in \cite{new}.

After inflation, but still in the high friction regime, the vector masses are sub-Planckian. However, the W and Z bosons are heavier than the tree-level cut off scale of the scalar-vector interactions $\sim \Lambda_M^3/(g^2\phi_0^2)$, as read off from the tree-level coupling in \eqref{gauge}. Here $\Lambda_M=(M^2 M_p)^{1/3}$ is the cut-off scale on a Minkowski background.

The assumption usually taken in the literature would be to consider the bosons to be decoupled from our effective ``low energy" theory. This might indeed be the easiest way to treat the gauge bosons.

Here we will also consider an additional approach. We will introduce a mechanism that brings the bosons masses below the vector-scalar cut-off, and in particular below the Hubble scale. In this way we can still treat the bosons as propagating degrees of freedom throughout the evolution of the Universe. We do so by introducing to our theory a non-minimal kinetic coupling of the gauge bosons to the Higgs. This serves to enhance the kinetic term of the vectors, thereby reducing both their mass and their effective coupling to gravity. We consider the interaction
\be\label{int34}
{\cal L}_\text{nm} &=&-\frac{1}{4} \left(g^{\alpha\mu}g^{\beta\nu} + M_p\frac{\xi^2 |{\cal H}|^2}{\Lambda_M^5}{}^{**}R^{\mu\nu\alpha\beta}\right)F_{\alpha\beta}F_{\mu\nu} ,
\ee
where ${}^{**}R^{\mu\nu\alpha\beta}$ is the double dual of the Riemann tensor and $\xi$ is a constant to be fixed by experiment.\footnote{Note that we have to couple through $^{**}R^{\mu\nu\alpha\beta}$ to render the coupling in the post-inflationary era negligible.}  
Note that this interaction is unique in the sense that it introduces neither ghosts, via higher derivatives, nor new degrees of freedom in the gauge-gravity sector \cite{loc}. 

By expanding into the graviton and vector, again in the gauge $\delta \phi=0$, one easily finds that the cut-off scale of this interaction is $\Lambda_H$, exactly the scale that we were able to gauge away in the scalar-scalar interactions, via a diagonalization of the graviton-Higgs system.

During the high friction regime, the gauge bosons will be canonically normalized as
\be
\bar A_{\mu} = \xi\phi\sqrt{\frac{\Lambda_H^3}{\Lambda_M^5}}A_\mu ,
\ee
with $\Lambda_H^3 = H^2 M_p$, changing their effective mass to
\be
m_\text{eff}^2 = \frac{3 g^2}{\xi^2}\Lambda_M^2\ .
\label{eq:fin_m}
\ee
It is easy to convince ourselves that $m_\text{eff}\ll \Lambda_H$ for any reasonable value for $\xi$, in particular $\xi= {\cal O}(1)$. Therefore, as announced, the coupling \eqref{int34} brings back the vectors to the low energy theory during inflation.

The interaction between the graviton and the vectors in the Higgs covariant derivative coupling \eqref{gauge} is instead suppressed by a scale $\sim \frac{\xi^2}{g^2}\frac{\Lambda_H^3}{\Lambda_M^2} \gg \Lambda_H$.

After the end of inflation, the vector masses reduce to the known values, $m^2 \sim g^2\phi^2$, while the strong coupling scale becomes $\Lambda_M$.

\section{Unitarity issues: post-inflation}
\label{sec:post-inf}

Although the New Higgs inflation might be completely unitary during inflation, one may ask whether an inflationary background can be consistently obtained from the Higgs boson, starting from a Minkowski background, without the necessity of integrating-in any new degree of freedom at low energies. 

Let us re-discuss the gauge choice for scalar fluctuations. Up to second order in the fluctuations, we have
\be\label{tru}
\phi(t+\delta t,x)=\phi_0+\delta\phi+\dot\phi_0\delta t+\delta\dot\phi\delta t+\frac{1}{2}\ddot\phi_0\delta t^2+\ldots\ .
\ee
Truncating this series at the first order would mean
\be
\dot \phi_0\delta t\gg \delta\dot\phi \delta t\ ,
\ee
or 
\be
\dot\phi_0\gg \delta\dot\phi\ .
\ee
When the time derivative of the scalar is too small, a linear truncation in \eqref{tru} will be inconsistent and so the expansion in $\delta t$ would no longer be useful.  In other words, at these low energies, the fluctuations of the field cannot be re-absorbed into a {\it linear} coordinate re-definition, i.e. into a longitudinal linear graviton. From a particle physics language, this just means that there is effectively no mixing between the graviton and the scalar in flat background.

The easier way to treat the system in this regime is thus to ''integrate out" gravity as in \cite{gal}. In that case, the non-minimal gravitational interaction appears as a non-trivial self-derivative coupling of the Higgs boson. The structure will be the one of a quartic Galileon \cite{galileon, gal}. 

Ignoring the gauge bosons and considering scales much larger than the electroweak, thus also ignoring $v$, we have that in the decoupling limit $M_p\rightarrow\infty$ but $\Lambda_M^3=M^2M_p<\infty$ \cite{gal}
\be\label{eq:gal}
{\cal L}_{\rm dec}=-\frac{1}{2}\partial_\mu\phi\partial^\mu\phi\left[1+\frac{(\square\phi)^2-\partial_{\mu\nu}\phi\partial^{\mu\nu}\phi}{2\Lambda_M^6}\right]-\frac{\lambda}{4}\phi^4\ .
\ee
The precise question is now whether we could create a background with Hamiltonian energy density larger than $\Lambda_M^4$.

Before answering this question, let us see why this {\it would not} be possible in a theory with a non-renormalizable potential. Let us take for example a potential $V=\frac{\phi^6}{\Lambda^2}$. The Hamiltonian of the system would be
\be
{\mathscr{H}}=\frac{\pi^2}{2}+\frac{1}{2}\partial_i\phi\partial^i\phi+\frac{\phi^6}{\Lambda^2}\ ,
\ee 
where the momentum is defined in the standard way to be $\pi\equiv \frac{\delta{\cal L}}{\delta\dot\phi}=\dot\phi$.

Suppose we want to have a large homogeneous background with ${\mathscr{H}}\gg \Lambda^4$, i.e. a background formed by a large number of particles with very large wavelength. This can only be realized by taking $\phi \gg \Lambda$. However, a quick inspection of quantum corrections reveals the (expected) inconsistency.
The one-loop correction to the effective potential, renormalized at the scale $\mu$, takes the form 
\be
V_\text{$1$-loop} \sim \frac{\phi^8}{\Lambda^4}\log{\frac{\phi}{\mu}} + \text{counter-terms} ,
\ee
The problem of large logarithms may be avoided by integrating the corresponding renormalization group equations. This, however, reveals the generation of additional couplings $\phi^8/\Lambda^4, ...$, with coefficients that are not generically small when going towards the UV.

Of course, this is nothing but the incarnation of the non-renormalizability of the potential; when interested in processes at high energies, the theory requires a measurement of infinitely many coupling constant and loses predictivity. Therefore, no statements can be made without specifying the UV completion of the theory.

Let us now instead consider our theory, a quartic galileon. After a lengthy but straightforward calculation, one obtains that
\be
{\mathscr{H}}=\frac{1}{2}\frac{\pi^2}{1+3\Delta}+\frac{1}{2}\partial_i\phi\partial^i\phi(1+\Delta)+\frac{\lambda\phi^4}{4}\ ,
\ee
where $\pi=\dot\phi(1+3\Delta)$ and 
\be
\Delta=\frac{1}{2\Lambda^6_M}\left[(\partial_i\partial^i\phi)^2-\partial_{ij}\phi\partial^{ij}\phi\right]\ .
\ee
In passing, we note that all Galilean theories are quadratic in momenta and, therefore, path integral integration of momenta can be readily performed. Thus, Galilean theories can be generically quantized in the Lagrangian path integral.

If we now consider a homogeneous field $\phi$ which is large in amplitude, $\phi\gg \Lambda_M$, but small in spatial momenta $\frac{\partial_i\partial^i\phi}{\phi}\ll\Lambda_M^2$, we would {\it not} encounter the above strong coupling problem. In other words, in this case, we would be allowed to consider a homogeneous background field formed by a large number of quanta and very large wavelength. This is precisely what we need for inflation, as also explained in \cite{giainf}. In other words, zero-momentum quantum corrections are under control also for large field values, as our potential is renormalizable.
This is due to the famous non-renormalization theorem of the Galilean couplings \cite{nonren1,nonren2}. Although the shift symmetry is broken by the potential, no contributions to the effective potential involving the scale $\Lambda_M$ can be generated at the one loop level. Any loop containing the Galilean vertex vanishes in the zero-momentum limit\footnote{A thorough analysis of this issue at higher loops is postponed for future work.}. This theory also has another interesting property: it may allow fluctuations (or scatterings) with center of mass energy larger than $\Lambda_M$ without encountering unitarity problems. In fact, as the theory is self-derivatively coupled it enters into the realm of classicalizing theories \cite{class}\footnote{If, instead, one insists on a Wilsonian UV-completion, the above analysis relies on the implicit, but reasonable, assumption that couplings to new degrees of freedom respect an approximate shift symmetry.}.

In the Wilsonian realm instead, one may be still worried about the role of possible higher dimensional operators emerging as effective field theory corrections to our (low-energy) theory. 

The crucial point here is again the fact that potentially dangerous higher dimensional operators correcting our low-energy theory will appear either in powers of $\frac{p^2}{\Lambda_M^2}$ due to the derivative nature of the Galilean interaction, or in the form of the usual Coleman-Weinberg contribution due to the renormalizable potential. The former, albeit certainly being very important for high-energy scatterings, will give negligible contributions in the zero momentum limit. 
Thus, predictions related to large field values with small momentum ($p^2\ll \Lambda_M^2$) will be independent upon these additional higher dimensional operators. Once again, this is precisely what we need to construct a Friedmann-Lema\^itre-Robertson-Walker gravitational background.

Finally, we would like to comment on the fact that, similarly to the gravitational progenitor, this theory has a background dependent cut-off. On an inhomogeneous background, quantum corrections to Galilean theories are suppressed by a scale that increases with increasing inhomogeneities \cite{nonren2, galoneloop, tetradis}. 

Whenever the energy of the potential energy overcomes the scale $(MM_p)^{1/2}$, where gravity will be re-integrated-in, the strong coupling scale of the system will start to grow with the homogeneous Friedmann background. As explained before, this is due to the non-trivial re-canonicalization of the Higgs boson.

\section{Conclusions}
The discovery of polarized B-modes in the CMB by BICEP2 has completely changed our prospective of inflation since the release of the Planck results. With this new data, if inflation ever occurred, it must be chaotic-type, i.e. the excursion of the canonically normalized inflaton must be trans-Planckian. With the discovery of the Higgs boson at the LHC, it is very tempting to seriously consider the minimal scenario in which the Higgs boson is the inflaton. Here we showed that this would be compatible with Planck and BICEP2 data if the Higgs boson is non-minimally kinetically coupled to curvatures, as in the New Higgs Inflationary scenario of \cite{new}. In particular, we show that the mild tension between Planck and BICEP2 data can be released by a negative running of the New Higgs inflation spectral index. 

Finally, we have argued that our model is unitary throughout the whole evolution of the Universe. In particular, in the original model of Higgs inflation the gauge sector can be considered decoupled from the low energy effective theory. 

A non-minimal interaction of the gauge bosons to the Higgs and the double-dual Riemann tensor will, on the other hand, significantly lower the gauge boson masses during inflation. In this case, we showed that the gauge bosons can be treated within our effective field theory even during inflation.

\acknowledgments
CG would like to thank Oscar Cata, Daniel Flassig and Alex Pritzel for very enlightening discussions. CG thanks Gia Dvali and Alex Kehagias for comments on the first version of the manuscript. CG and NW are supported by the Humboldt foundation. YW thanks Jun'ichi Yokoyama for comments on the first draft of our manuscript. YW is supported by Grant-in-Aids for Scientific Research on Innovative Areas No.~21111006, Scientific Research No.~23340058 and JSPS Fellow No.~269337.


\begin{thebibliography}{99}

\bibitem{lhc}
  S.~Chatrchyan {\it et al.}  [CMS Collaboration],
  Phys.\ Lett.\ B {\bf 716} (2012) 30
  [arXiv:1207.7235 [hep-ex]].
  G.~Aad {\it et al.}  [ATLAS Collaboration],
  Phys.\ Lett.\ B {\bf 716} (2012) 1
  [arXiv:1207.7214 [hep-ex]].

\bibitem{planck} 
 P.~A.~R.~Ade {\it et al.}  [Planck Collaboration],
  arXiv:1303.5082 [astro-ph.CO].
  
\bibitem{bicep}
P.~A.~R.~Ade {\it et al.}  [BICEP2 Collaboration],
  arXiv:1403.3985 [astro-ph.CO].

\bibitem{linde}
  A.~D.~Linde, 
  Phys. Lett. B 129, 177
(1983)

\bibitem{linden}
A.~D.~Linde,
  Contemp.\ Concepts Phys.\  {\bf 5} (1990) 1
  [hep-th/0503203].

\bibitem{riottoke}
A.~Kehagias and A.~Riotto,
  arXiv:1403.4811 [astro-ph.CO].

\bibitem{mukhanov}
V.~Mukhanov,
  Cambridge, UK: Univ. Pr. (2005) 421 p
  
\bibitem{running}
G.~Degrassi, S.~Di Vita, J.~Elias-Miro, J.~R.~Espinosa, G.~F.~Giudice, G.~Isidori and A.~Strumia,
  JHEP {\bf 1208} (2012) 098
  [arXiv:1205.6497 [hep-ph]].
  
\bibitem{riotto}
A.~Kehagias, A.~M.~Dizgah and A.~Riotto,
  Phys.\ Rev.\ D {\bf 89} (2014) 043527
  [arXiv:1312.1155 [hep-th]].

\bibitem{shaposhnikov}
F.~L.~Bezrukov and M.~Shaposhnikov,
  Phys.\ Lett.\ B {\bf 659} (2008) 703
  [arXiv:0710.3755 [hep-th]].
  
\bibitem{fumi}
K.~Nakayama and F.~Takahashi,
  arXiv:1403.4132 [hep-ph].

\bibitem{new}
C.~Germani and A.~Kehagias,
  Phys.\ Rev.\ Lett.\  {\bf 105} (2010) 011302
  [arXiv:1003.2635 [hep-ph]].
  
\bibitem{fotis}
F.~Farakos, C.~Germani, A.~Kehagias and E.~N.~Saridakis,
  JHEP {\bf 1205} (2012) 050
  [arXiv:1202.3780 [hep-th]].
  
\bibitem{fotis2}
I.~Dalianis and F.~Farakos,
  arXiv:1403.3053 [hep-th].
  
\bibitem{alexpert}
C.~Germani and A.~Kehagias,
  JCAP {\bf 1005} (2010) 019
   [Erratum-ibid.\  {\bf 1006} (2010) E01]
  [arXiv:1003.4285 [astro-ph.CO]].

\bibitem{yuki}
 C.~Germani and Y.~Watanabe,
  JCAP {\bf 1107} (2011) 031
   [Addendum-ibid.\  {\bf 1107} (2011) A01]
  [arXiv:1106.0502 [astro-ph.CO]].
  
\bibitem{somehow}
  C.~Germani,
  Rom.\ J.\ Phys.\  {\bf 57} (2012) 841
  [arXiv:1112.1083 [astro-ph.CO]].

\bibitem{wmap}
E.~Komatsu {\it et al.}  [WMAP Collaboration],
  Astrophys.\ J.\ Suppl.\  {\bf 192} (2011) 18
  [arXiv:1001.4538 [astro-ph.CO]].
  
\bibitem{Kobayashi:2011nu} 
  T.~Kobayashi, M.~Yamaguchi and J.~Yokoyama,
  Prog.\ Theor.\ Phys.\  {\bf 126}, 511 (2011)
  [arXiv:1105.5723 [hep-th]].

\bibitem{Kamada} 
  K.~Kamada, T.~Kobayashi, T.~Takahashi, M.~Yamaguchi and J.~Yokoyama,
  Phys.\ Rev.\ D {\bf 86}, 023504 (2012)
  [arXiv:1203.4059 [hep-ph]];
  K.~Kamada, T.~Kobayashi, T.~Kunimitsu, M.~Yamaguchi and J.~Yokoyama,
  Phys.\ Rev.\ D {\bf 88}, 123518 (2013)
  [arXiv:1309.7410 [hep-ph]];
  K. Kamada, T. Kobayashi, M. Yamaguchi and J. Yokoyama, Phys. Rev. D 83, 083515 (2011).
  
\bibitem{Tsujikawa} 
  S.~Tsujikawa,
  Phys.\ Rev.\ D {\bf 85}, 083518 (2012)
  [arXiv:1201.5926 [astro-ph.CO]].

\bibitem{maldacena}
J.~M.~Maldacena,
  JHEP {\bf 0305} (2003) 013
  [astro-ph/0210603].

\bibitem{sarah}
G.~Dvali, S.~Folkerts and C.~Germani,
  Phys.\ Rev.\ D {\bf 84} (2011) 024039
  [arXiv:1006.0984 [hep-th]].

\bibitem{luca}
  C.~Germani, L.~Martucci and P.~Moyassari,
  Phys.\ Rev.\ D {\bf 85} (2012) 103501
  [arXiv:1108.1406 [hep-th]].

\bibitem{loc}
C.~Germani,
  Phys.\ Rev.\ D {\bf 85} (2012) 055025
  [arXiv:1109.3718 [hep-ph]].
  
\bibitem{gal}
C.~Germani,
  Phys.\ Rev.\ D {\bf 86} (2012) 104032
  [arXiv:1207.6414 [hep-th]].

\bibitem{galileon}
  A.~Nicolis, R.~Rattazzi and E.~Trincherini,
  Phys.\ Rev.\ D {\bf 79} (2009) 064036
  [arXiv:0811.2197 [hep-th]].

\bibitem{nonren1}
  M.~A.~Luty, M.~Porrati and R.~Rattazzi,
  JHEP {\bf 0309} (2003) 029
  [hep-th/0303116].

\bibitem{nonren2}
  A.~Nicolis and R.~Rattazzi,
  JHEP {\bf 0406} (2004) 059
  [hep-th/0404159].
\bibitem{giainf}
G.~Dvali and C.~Gomez,
  arXiv:1312.4795 [hep-th].
  
\bibitem{class}
G.~Dvali, G.~F.~Giudice, C.~Gomez and A.~Kehagias,
  JHEP {\bf 1108} (2011) 108
  [arXiv:1010.1415 [hep-ph]].
  
\bibitem{tetradis}
  N.~Brouzakis, A.~Codello, N.~Tetradis and O.~Zanusso,
  arXiv:1310.0187 [hep-th].
  N.~Brouzakis and N.~Tetradis,
  arXiv:1401.2775 [hep-th].
  
\bibitem{galoneloop}
  K.~Hinterbichler, M.~Trodden and D.~Wesley,
  Phys.\ Rev.\ D {\bf 82} (2010) 124018  [arXiv:1008.1305 [hep-th]].
  C.~de Rham, G.~Gabadadze, L.~Heisenberg and D.~Pirtskhalava,
  Phys.\ Rev.\ D {\bf 83} (2011) 103516
  [arXiv:1010.1780 [hep-th]].
  T.~de Paula Netto and I.~L.~Shapiro,
Phys.\ Lett.\ B {\bf 716} (2012) 454  [arXiv:1207.0534 [hep-th]].  
  A.~Codello, N.~Tetradis and O.~Zanusso,
JHEP {\bf 1304} (2013) 036  [arXiv:1212.4073 [hep-th]]; 
  C.~de Rham, G.~Gabadadze, L.~Heisenberg and D.~Pirtskhalava,
  Phys.\ Rev.\ D {\bf 87} (2013) 8,  085017
  [arXiv:1212.4128].
  C.~de Rham, L.~Heisenberg and R.~H.~Ribeiro,
  Phys.\ Rev.\ D {\bf 88} (2013) 084058
  [arXiv:1307.7169 [hep-th]].
  
  
  
  
\end{thebibliography}
\end{document}